\documentclass[aps,prd,amsmath,amssymb,twocolumn,preprintnumbers]{revtex4}
\pdfoutput=1
\usepackage{graphicx}
\usepackage{color}
\newcommand{\beq}{\begin{eqnarray}}
\newcommand{\eeq}{\end{eqnarray}}

\newcommand{\bmp}{\noindent\begin{minipage}{16cm}}
\newcommand{\emp}{\end{minipage}\vskip 7mm} 

\usepackage{bm}
\usepackage{bbm}
\usepackage{pxfonts}

\usepackage[ margin=5pt, font=normalsize,labelfont=bf,justification=raggedright]{caption}
\usepackage{subfig}
\usepackage{youngtab}
\usepackage{slashed}

\definecolor{rossoCP3}{cmyk}{0,.88,.77,.40}

\usepackage{fancyhdr}
\begin{document}
\title{\Large  \color{rossoCP3} Hot Conformal Gauge Theories} 
\author{Matin {\sc Mojaza}$^{\color{rossoCP3}{\varheartsuit}}$}
\email{mojaza@cp3.sdu.dk} 
\author{Claudio {\sc Pica}$^{\color{rossoCP3}{\varheartsuit}}$}
\email{pica@cp3.sdu.dk} 
\author{Francesco {\sc Sannino}$^{\color{rossoCP3}{\varheartsuit}}$}
\email{sannino@cp3.sdu.dk} 
\affiliation{
$^{\color{rossoCP3}{\varheartsuit}}${ CP}$^{ \bf 3}${-Origins}, 
University of Southern Denmark,  Campusvej 55, DK-5230 Odense M, Denmark.}
\begin{abstract}
We compute the nonzero temperature free energy up to the order $g^6 \ln(1/g)$ in the 
coupling constant for vector like $SU(N)$ gauge theories
featuring matter transforming according to different representations of the underlying gauge group.  The number of matter fields, i.e. flavors, is arranged in such a way that the theory develops a perturbative stable infrared fixed point at zero temperature. 
Due to large distance conformality we trade the coupling constant with its fixed point value and define a reduced free energy which depends only on the number of flavors, colors and matter representation. 

We show that the reduced free energy changes sign, at the second, fifth and sixth order in the coupling, when decreasing the number of flavors from the upper end of the conformal window. If the change in sign is interpreted as signal of an instability of the system then we infer a critical number of flavors. Surprisingly this number, if computed to the order $g^2$, agrees with previous predictions for the lower boundary of the conformal window for nonsupersymmetric gauge theories. The higher order results tend to predict a higher number of critical flavors. These are universal properties, i.e. they are independent on the specific matter representation. 
 \\[.1cm]
{\footnotesize  \it Preprint: CP$^3$-Origins-2010-45}
\end{abstract}

\maketitle
\thispagestyle{fancy}

%
\section{Introduction}

Non-Abelian gauge theories are expected to exist in a number of different phases which can be classified according to the force measured between two static sources.
The knowledge of this phase diagram is relevant 
for the construction of extensions of the 
Standard Model (SM) that invoke dynamical electroweak symmetry
breaking \cite{Weinberg:1979bn,Susskind:1978ms}.   An up-to-date review is \cite{Sannino:2009za} while earlier reviews are \cite{Hill:2002ap,Yamawaki:1996vr}.  The phase diagram is also useful in providing ultraviolet completions of 
unparticle \cite{Georgi:2007ek} models \cite{Sannino:2008nv,Sannino:2008ha} and it has been investigated recently using different analytical methods \cite{Sannino:2004qp,Dietrich:2006cm,Ryttov:2007sr,Ryttov:2007cx,Sannino:2009aw,Gies:2005as,Poppitz:2009uq,Antipin:2009wr,Jarvinen:2009fe,Fukano:2010yv,Frandsen:2010ej}.

Here we wish to understand, in a rigorous way, the dynamics of gauge theories lying in the conformal window at nonzero temperature. The physical applications are numerous ranging from the above mentioned models of dynamical electroweak symmetry breaking to cosmology \cite{Cline:2008hr,Jarvinen:2009wr,Jarvinen:2009pk,Jarvinen:2009mh,Jarvinen:2010ks}.  Thermodynamical properties of these gauge theories were also investigated in the literature using holographic models \cite{Jarvinen:2009fe,Alanen:2009na,Alanen:2010tg}. 

Our starting point are asymptotically free vector-like gauge theories near the Banks-Zaks infrared stable fixed point \cite{Banks:1981nn}. The presence of such a perturbative fixed point allows a controllable computation of the free energy for these theories which we carry till order $g^6$ in the gauge coupling. In fact we expect perturbation theory to work, in principle, for any temperature in contrast to the case of a {\it confining} theory for which perturbation theory is limited to asymptotically large temperatures compared to its intrinsic scale.  The absence of an intrinsic scale in the theory is evident in having a free energy directly proportional to the fourth power of the temperature, for any temperature. We will trade the value of the coupling with its value at the infrared fixed point turning the coefficient of $T^4$ into an algebraic expression of the number of flavors, colors and matter representation, encoding a great deal of information of the underlying gauge theory.

Despite the fact that we can keep the coupling constant small it is a fact that perturbation theory breaks down, at finite temperature, due to the loss of analyticity in the coupling associated to the presence of infrared singularities.
For the free energy this problem sets in at $\mathcal{O} (g^6)$, or four-loop order\cite{Linde:1980ts, Gross:1980br}. 
At best one can assume that the free energy is computable to $\mathcal{O} (g^6)$, though not via loop diagrams.
The highest  order one can achieve using Feynman diagrams is $\mathcal{O}\left (g^6\ln(1/g)\right )$ 
 and  was recently determined in \cite{Kajantie:2002wa}.  We adapt their results for the case of gauge theories featuring large distance conformality while generalizing the discussion to any matter representation, different number of colors and flavors.

We discover a number of surprising features when plotting the free energy, for a given matter representation, as function of the number of flavors: i) There is a change in sign of the free energy at a critical number of flavors whose value depends on the representation it belongs and the order to which the computations were carried, ii)This number is smaller than the one for which asymptotic freedom is lost. 

 It is tempting to identify it with the critical number of flavors below which, at zero temperature, conformality is lost. The obvious caveat is that as we decrease the number of flavors away from the point when asymptotic freedom is lost perturbation theory ceases to be reliable and therefore we interpret this phenomenon only as {\it strong} indication that the finite temperature free energy is aware of the nontrivial underlying gauge dynamics.

Another amusing feature is that at the two-loops level, when the results are scheme independent, the change in sign of the free energy occurs for a given number of flavors which is surprisingly close to the one predicted using the Schwinger-Dyson results \cite{Dietrich:2006cm} as well as the Ryttov-Sannino $\beta$ function \cite{Ryttov:2007cx}. This value becomes larger when going to the fifth and sixth order in the coupling. 

\section{Review of the Free energy computation}
The perturbative free energy at finite temperatures was
computed to order $\mathcal{O}\left(g^6\ln(1/g)\right)$ in \cite{Kajantie:2002wa} and in the $\overline{\text{MS}}$ scheme.
The result was derived
using an effective field theory approach utilizing
matching of coefficients in the effective theory expression with
the dimensionally regularized perturbative expansion.
An in depth presentation of the method can be found in \cite{Braaten:1995jr}, where the order $\mathcal{O}(g^5)$ was
recomputed. In order to present the results in a relatively self-contained way we briefly review  the method here.

The perturbative free energy for a
massless asymptotically free gauge theory receives contributions from the
following three mass scales: $2 \pi T$, $g T$ and $g^2 T$. They are respectively, the particle momentum in the plasma, the onset of the color-electric (Debye) screening and finally the onset
of color-magnetic screening.  The idea is then to construct an effective field theory
that reproduce static observables at the different scales.
This is done by the method of dimensional reduction, where
the static properties of a $3+1$-dimensional field theory
at high temperatures are expressed in terms of an effective 
field theory in $3$ space dimensions\cite{Gross:1980br, Appelquist:1981vg}.

The free energy density expressing the static equilibrium properties
of the plasma is given by the usual logarithm of the partition
function:
\begin{align}
F &= - \frac{T}{V} \ln \mathcal{Z},\\
\mathcal{Z} &= \int \mathcal{D} A_\mu  \mathcal{D}\psi  \mathcal{D}\bar\psi \exp\left(-\int_0^\beta d\tau \int d^dx \mathcal{L}\right), \\
\mathcal{L} &= \frac{1}{4} F_{\mu \nu}^a F^{a \mu \nu} + \bar \psi \slashed D \psi.
\end{align}
As it is customary to introduce finite temperature by euclideanizing the time dimension $t = -i \tau$ and compactifying it with a period $\beta  = 1/T$, with $T$ the temperature,  and the bosonic (fermionic) fields respecting periodic (antiperiodic) boundary conditions.  In regimes when the gauge coupling, $g$ is small the free energy density can be computed perturbatively.

On the other hand, by dimensional reduction one can
compute the free energy density using an effective field theory in
$3$ space dimensions, by seperating the electro- and magnetostatic
parts of the Lagrangian. The free energy density is expressed
as
\begin{align}
F = T \left [ f_E(T,g;\Lambda_E) \right . &+ f_M(m_E^2,g_E,\lambda_E^{(i)},\ldots; \Lambda_E, \Lambda_M) \nonumber\\
 &\left. + f_G(g_M,\ldots; \Lambda_M) \right ],
\end{align}
where the effective free energy densities $f_E, f_M$ and $f_G$ represents
the contributions from the three scales, i.e. $f_E$
gives the contribution from the momentum scale by effectively integrating
out the fermions and the high momentum degrees of freedom down to the scale $\Lambda_E$ 
corresponding to a distance of order $1/(gT)$. At greater distances the
fields are replaced by electrostatic and magnetostatic gauge fields $A_0^a(\mathbf x)$
and $A_i^a(\mathbf x)$, which are proportional to 
the zero-frequency modes of the gauge fields $A_\mu(\tau, \mathbf{x})$
that are the only fields able to propagate over such distances\cite{Braaten:1994na}.
$f_E$ is then the normalization for this transition, i.e.
\begin{align}
\mathcal{Z} = e^{-f_E(\Lambda_E)V}\int^{\Lambda_E} \mathcal{D}A_\mu^a(\mathbf{x}) \exp \left ( - \int d^d x \mathcal{L}_E\right),
\end{align}
where the Lagrangian is now an
effective electrostatic Lagrangian:
\begin{align}
\mathcal{L}_E = \frac{1}{2} \text{Tr} {F_{ij}}^2 &+ \text{Tr} [D_i,A_0]^2 + m_E^2 \text{Tr} A_0^2\nonumber \\
&+ \lambda_E^{(1)} \left[\text{Tr}(A_0^2)\right]^2 + \lambda_E^{(2)} \text{Tr}A_0^4 + \delta \mathcal{L}_E.
\end{align}
$\delta \mathcal{L}_E$ represents higher order interaction terms which contributes
beyond the order $\mathcal{O}(g^6)$. Shorthand notation has been used
for the gauge fields, i.e $A_\mu = T^a A_\mu^a$ and the corresponding
 gauge coupling is denoted  by $g_E$.
This Lagrangian defines $f_M$ which is dependent on the still lower
momentum scale $\Lambda_M$, corresponding to the distance $1/(g^2T)$.
Again at greater distances, only the magnetostatic gauge fields play a role:
\begin{align*}
\mathcal{Z} = e^{-f_E(\Lambda_E)V}e^{-f_M(\Lambda_E, \Lambda_M) V}
\int^{\Lambda_M} \mathcal{D}A_i^a(\mathbf{x}) \exp \left ( - \int d^d x \mathcal{L}_M\right),
\end{align*}
with the effective magnetostatic Lagrangian
\begin{align}\label{eq:LM}
\mathcal{L}_M = \frac{1}{2} \text{Tr}(F_{ij}^2) + \delta \mathcal{L}_M,
\end{align}
containing the gauge coupling $g_M$.
Note that the direct identification of the normalization 
functions $f_E$ and $f_M$ with the free energy densities
of the respective Lagrangians is strictly true only when 
using dimensional regularization to cut off the
ultraviolet and infrared divergences in the perturbation
expansions. In the same sense, $f_G$ is identified with the 
integral expression in the above partition function.

 The Lagrangian \eqref{eq:LM} defines a confining theory 
 when the higher order terms are not considered and
is thus non-perturbative. However, $f_G$ can 
still be expressed as a power series in $g$ \cite{Braaten:1994na}.
The leading order is proportional to  $(g^2 T)^3$ and
the coefficient can be determined by lattice computations only.
However, the logarithmic ultraviolet divergence coming 
from the scale $\Lambda_M$ can be evaluated exactly
by matching the coefficient with that in $f_M$, since 
the expression must be scale invariant.
In this way one proceeds backwards and matches all
coefficient with appropriate tuning to get rid of the somewhat
arbitrary momentum scales. Finally one
ends up with an exact expression for the free energy to order
$\mathcal{O}\left ( g^6\ln(1/g) \right)$, while the order $\mathcal{O}(g^6)$
coefficient remains unknown and uncomputable from perturbation methods.
One also shifts the scale of the coupling constant
from the dimensional regularization scale to
an arbitrary renormalization scale $\mu$ by
using the renormalization group equation for the running 
of the coupling constant.

 We here express 
the leading order magnitudes only and refer to \cite{Kajantie:2002wa} for the full result.
\begin{align*}
&f_E \sim T^4, \quad m_E^2 \sim g^2T^2, \quad g_E^2 \sim g^2 T, \\
&\lambda_E^{(1)} \sim g^4 T,\quad  \lambda_E^{(2)} \sim g^4 T,
\quad g_M^2 \sim g^2 T.
\end{align*}
Note that the perturbation expansion parameter in $\mathcal{L}_E$
is $g_E^2/m_E$ which is of order $g$. Thus the
perturbation expansion by this method is an expansion in $g$ rather
than $g^2$.

\section{Hot Conformal Free Energy @ $\mathbf{\cal O}(g^2)$}\label{sec:g2}
Our starting point is a generic asymptotically free gauge theory with $N_f$ Dirac flavors transforming according to the representation $r$ of the underlying gauge group. We will consider, to this order, also the case of ${\cal N}=1$ supersymmetric gauge theories for reasons which will become clearer shortly. 

To be more specific, throughout this paper we will consider
matter transforming according to four different but single representations, i.e.
the adjoint representation (denoted $G$) under the
gauge group $SU(2)$, 
the fundamental representation under
the gauge groups $SU(3)$ and $SU(2)$, the 
two-index symmetric representation under the
gauge group $SU(3)$ 
and the two-index antisymmetric representation
under the gauge group $SU(4)$. 

The relevant group normalization factors are:
\begin{equation}
\text{Tr}[ T^a_r T^b_r ]  = T[r] \delta^{ab}, \qquad 
T^a_r T^a_r = C_2[r] \mathbf{1},
\end{equation}
where $T^a_r$ is the $a$-th group generator in the representation $r$ and  $a=1,\dots, d[G]$. We denote with $d[r]$ the dimension of the representation. $T[r]$ and $C_2[r]$ are related via the identity $C_2[r] d[r] = T[r] d[G]$.
In table \ref{groupfactors}, for completeness, we list the normalization used 
for the group factors in the different representations. We list in the last column also
the number of colors which will be considered. The normalizations were taken from \cite{Ryttov:2007cx}.
\begin{table}[!htbp]
\caption{Normalization of the relevant group factors for the representations used throughout this paper.}
\label{groupfactors}
\begin{tabular}{c | ccc | c}
$r$ 		& $T[r]$	& $C_2[r]$ & $d[r]$& $N$  \\[1mm]
\hline\\[-3mm]
$G$  	& $N$ 	& $N$ 	& $N^2-1$  & $2$\\[1mm]
$\tiny\yng(1)$ 	& $\tiny\frac{1}{2}$ & $\tiny\frac{N^2-1}{2N}$ & $N$ & $2,3$\\[1mm]
$\tiny\yng(2)$ & $\tiny\frac{N+2}{2}$ & $\tiny\frac{(N-1)(N+2)}{N}$ & $\tiny\frac{N(N+1)}{2}$  & $3$\\[1mm]
$\tiny\yng(1,1)$ & $\tiny\frac{N-2}{2}$ & $\tiny\frac{(N+1)(N-2)}{N}$ & $\tiny\frac{N(N-1)}{2}$ & $4$
\end{tabular}
\end{table}

The $\beta$ function up to four-loop order 
 \begin{align}
 \beta (g) = - \frac{\beta_0}{(4\pi)^2} g^3 - \frac{\beta_1}{(4\pi)^4} g^5
 -\frac{\beta_2}{(4\pi)^6} g^7 - \frac{\beta_3}{(4\pi)^8} g^9 + \mathcal{O}(g^{11}),
 \end{align}
was computed in \cite{vanRitbergen:1997va}. As for the free energy expression the four-loop
$\beta$ function  is also 
computed in the $\overline{\text{MS}}$ scheme, thus
no ambiguities in the scheme-dependence of the expressions arise. 
 Only $\beta_0$ and $\beta_1$ are scheme-independent and read: 
 \begin{align}
\beta_0 &= \frac{11}{3} C_2[G] - \frac{4}{3} T[r] N_f, \\
\beta_1 &= \frac{34}{3} C_2^2[G] - \left ( \frac{20}{3} C_2[G] + 4 C_2[r] \right ) T[r] N_f.
\end{align}
Asymptotic freedom is lost when the lowest order coefficient, $\beta_0$
changes sign. This occurs at
\begin{align}
N_f^{\rm AF} := \frac{11}{4} \frac{C_2[G]}{T[r]}.
\end{align}
{}For a given fermion representation, the second coefficient, $\beta_1$ is negative below this
critical number of flavors and an infrared-stable fixed point develop which is known as the Banks-Zaks fixed point\cite{Banks:1981nn}.
Such a theory display large distance conformality.  

The Banks-Zanks fixed point disappears when $\beta_1$ changes sign. This occurs at: 
\begin{align}
\label{eq:NfIII}
N_f^{\rm Lost} := \frac{17 C_2[G]}{10 C_2[G] + 6 C_2[r]} \frac{C_2[G]}{T[r]}.
\end{align}

We are now equipped to investigate the {\it conformal} free energy by starting with the nonsupersymmetric case to the order $g^2$:
\begin{align}
\frac{F}{\pi^2 T^4 } = - \frac{d[G]}{9} \left [ \frac{1}{5} \right. &+ \frac{7}{20}\frac{d[r]}{d[G]}N_f \nonumber \\ 
&-  \left. \left ( C_2[G] + \frac{5}{2}T[r] N_f \right ) \frac{g^2(\mu)}{(4\pi)^2} \right ].
\end{align}

For the supersymmetric case we have, 
\begin{align}
 \beta_0^{\rm SUSY} & = 3 C_2[G] - 2 T[r] N_f \ , \\[2mm]
 \beta_1^{\rm SUSY} & = 6 C_2^2[G] - 4(C_2[G] + 2C_2[r]) T[r] N_f \ ,
 \end{align}
leading to:
 \begin{align}
 N_{f,SUSY}^{\rm AF} &= \frac{3}{2}\frac{C_2[G]}{T[r]}, \\[2mm]
 N_{f,SUSY}^{\rm Lost} & =  \frac{3 C_2[G]}{2C_2[G] + 4 C_2[r]}\frac{C_2[G]}{T[r]}.
\end{align}
We obtain for the supersymmetric free energy\cite{Grundberg:1995cu}:
\begin{align}
\frac{F_{SUSY}}{\pi^2 T^4} =- \frac{d[G]}{24} \left [ 1 \frac{}{} \right.& + 2\frac{d[r]}{d[G]}N_f \nonumber \\ 
&- \left. 6\left ( C_2[G] + 6T[r] N_f \right ) \frac{g^2(\mu)}{(4\pi)^2} \right ].
\end{align}
To determine the free energy dependence on the number of flavors and colors in the perturbative regime of the conformal window we replace the coupling constant with the Banks-Zaks fixed point value $g^*$
at two-loop order, given in appendix \ref{IRFP}. 
The free energies read:
\begin{align*}
\frac{F^*}{\pi^2 T^4 } = 
&- \frac{d[G]}{9} \left [ \frac{1}{5} + \frac{7}{20}\frac{d[r]}{d[G]}N_f \right. \nonumber \\ 
&+ \left. \frac{\left ( C_2[G] + \frac{5}{2}T[r] N_f \right ) \left ( 11 C_2[G] - 4 T[r] N_f \right)}{
34 C_2^2[G] - ( 20 C_2[G] + 12 C_2[r] ) T[r] N_f} \right ],\\
 \frac{F_{SUSY}^*}{\pi^2 T^4} = &- \frac{d[G]}{24} \left [ 1 + 2\frac{d[r]}{d[G]}N_f \right. \nonumber \\ 
&+ \left. \frac{3 \left ( C_2[G] + 6T[r] N_f \right ) \left ( 3 C_2[G] - 2 T[r] N_f \right)}{
3 C_2^2[G] - 2(  C_2[G] + 2 C_2[r] ) T[r] N_f} \right ].
\end{align*}
We observe immediately that due to the {\it conformal} large distance nature of our theories the free energy dependence on the energy scale is only via the temperature which factors out leaving behind, as expected, a numerical factor containing information on the specific theory studied. These coefficients are {\it universal}, i.e. independent on renormalization schemes. 

\begin{figure*}[t]
	\subfloat[$SU(3)$ with fundamental fermions.]{\label{subfig:3fund} \includegraphics[width=0.25\textwidth]{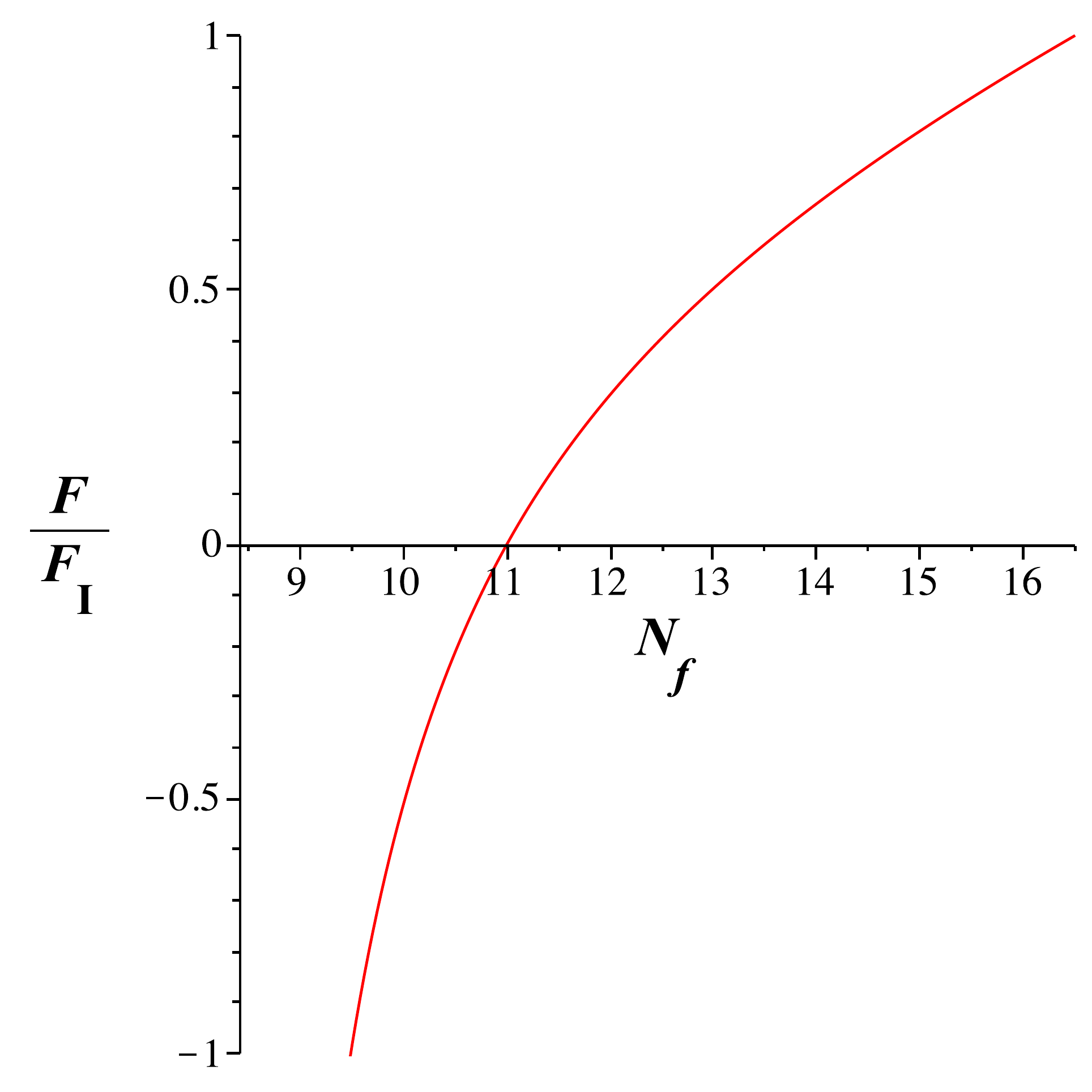}}
	\subfloat[$SU(2)$ with fundamental fermions.]{\label{subfig:2fund} \includegraphics[width=0.25\textwidth]{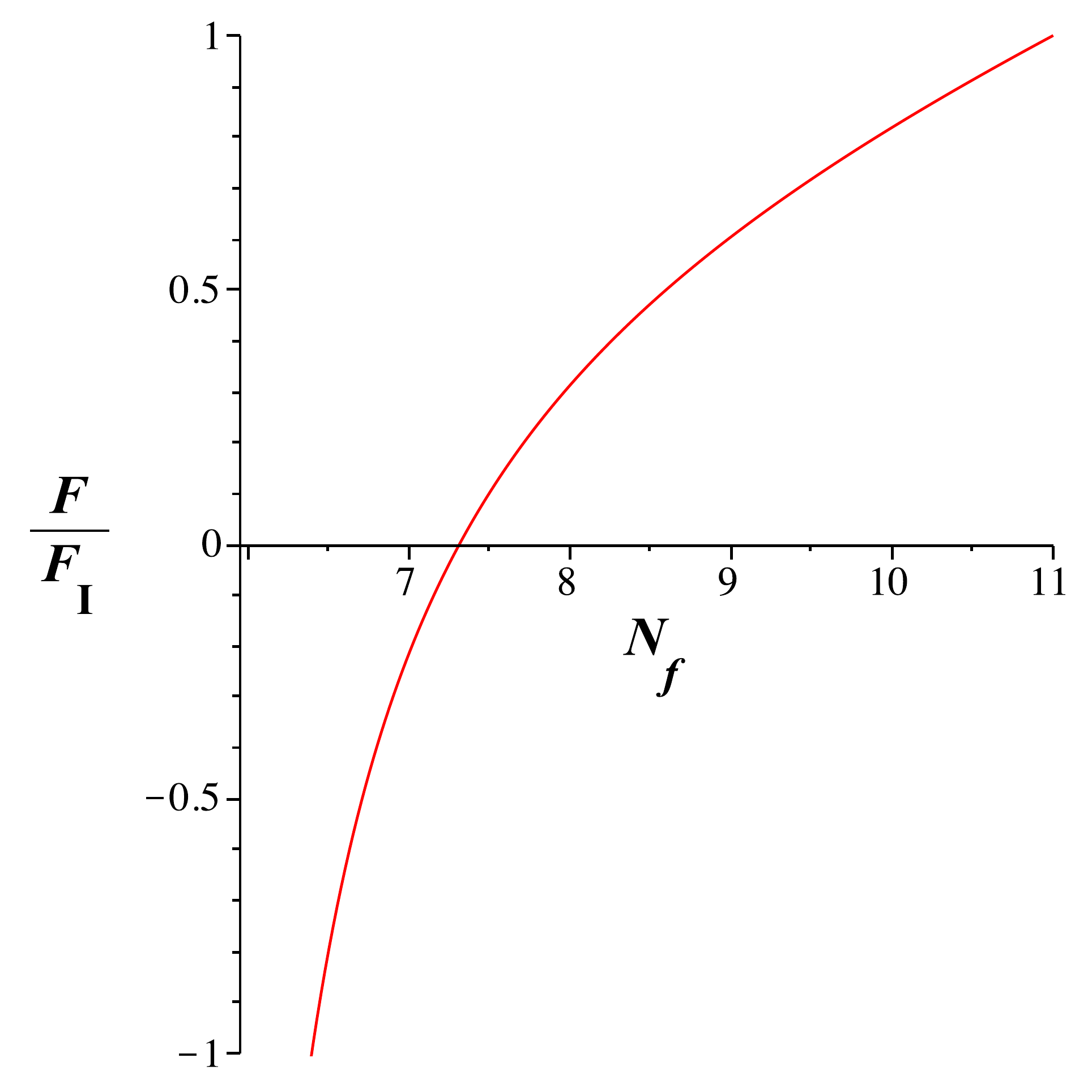}}
	\subfloat[$SU(3)$ with two-index symmetric fermions.]{\label{subfig:3sym} \includegraphics[width=0.25\textwidth]{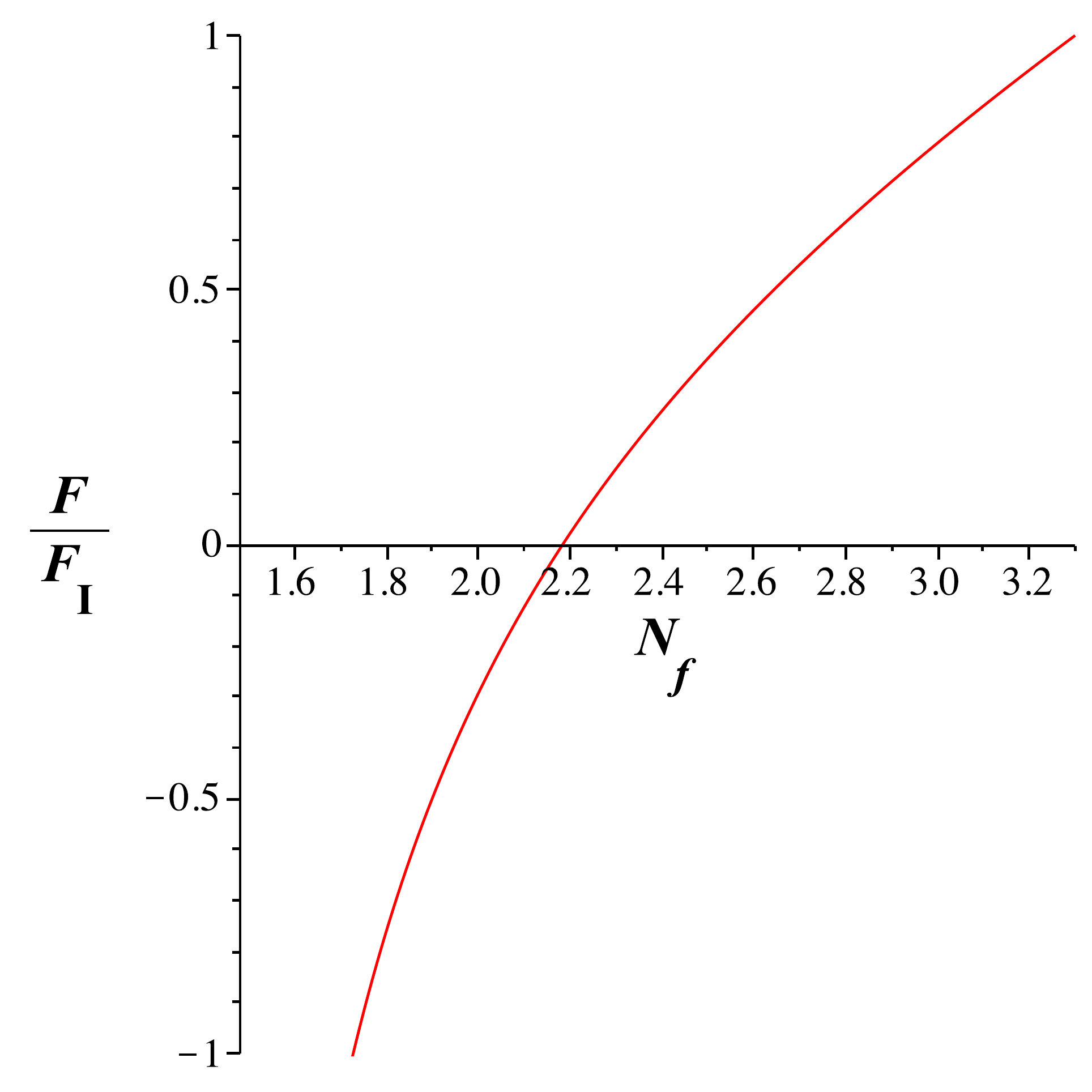}}
	\\	
	\subfloat[$SU(4)$ with 2-index antisymmetric fermions.]{\label{subfig:4antisym} \includegraphics[width=0.25\textwidth]{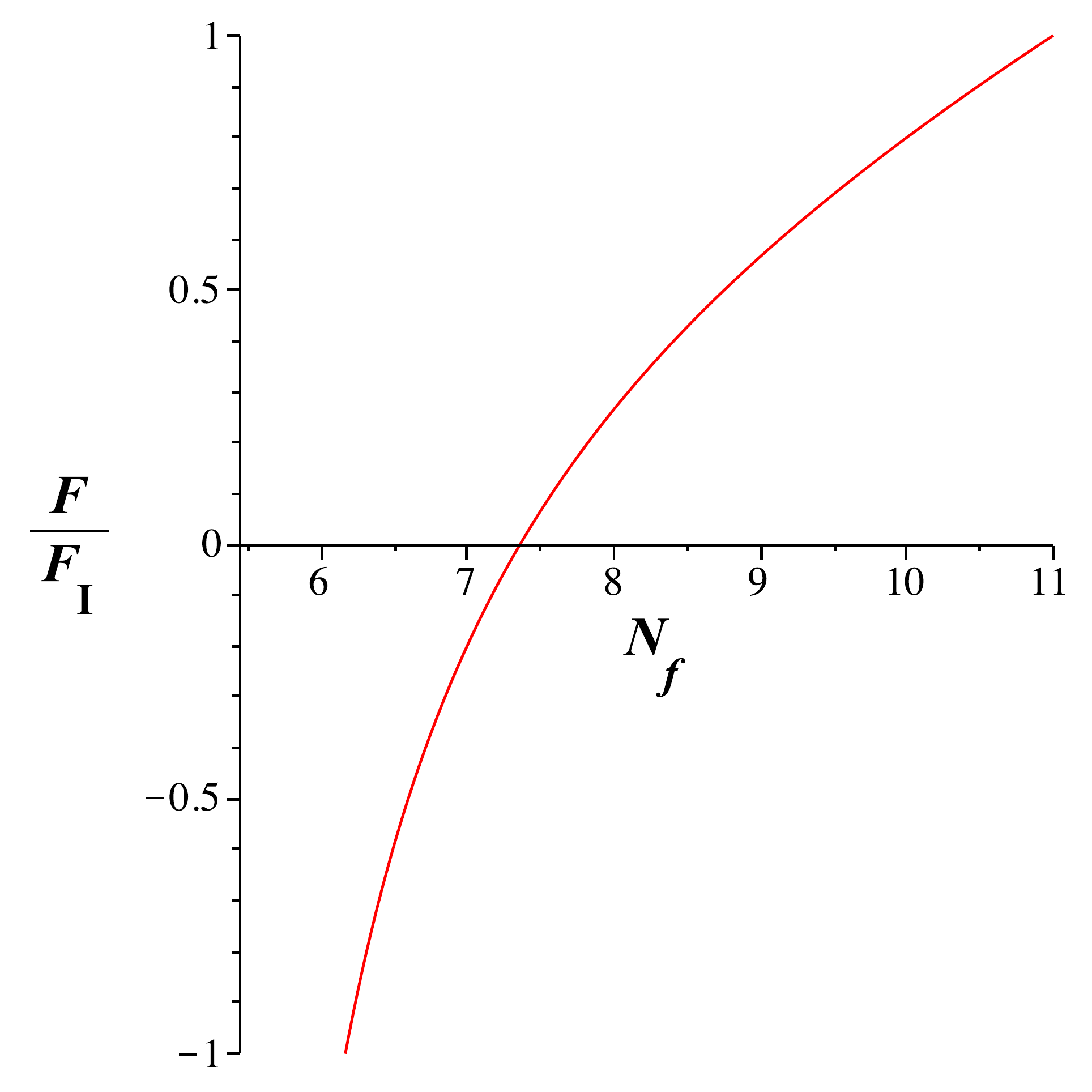}}
	\subfloat[$SU(2)$ with adjoint fermions.]{\label{subfig:2adj} \includegraphics[width=0.25\textwidth]{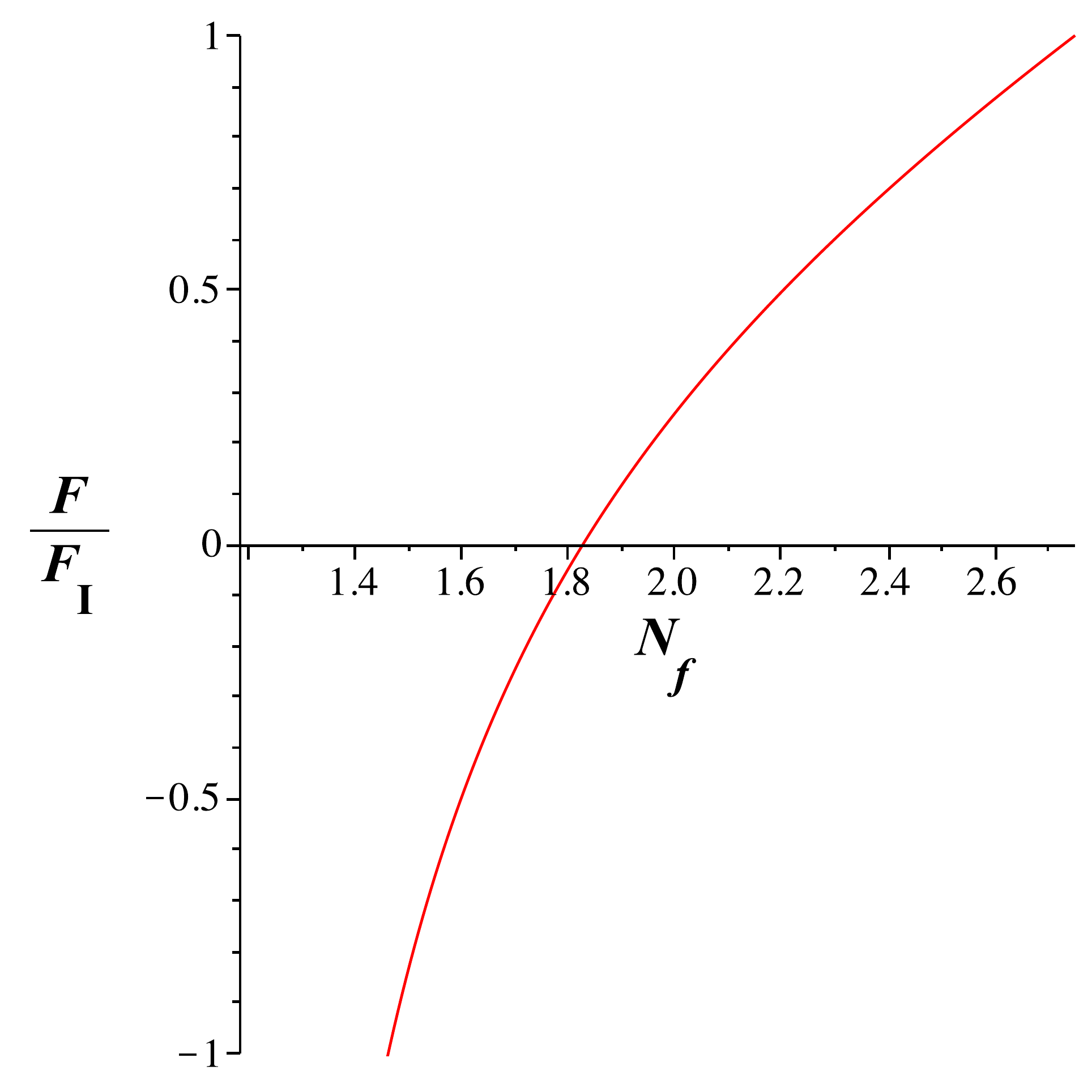}}
	\caption{Normalized free energy in the conformal window for different theories.}
	\label{fig:plots}
\end{figure*}

We can now plot the free energies as function of number of flavors for different number of colors and gauge theories. We choose to normalize our results to the free energy obtained, at order $g^2$,  when replacing the number of flavors with the one for which asymptotic freedom is lost for any given underlying gauge theory (defined as $F_I$ in the plots).  
The results are shown in figure \ref{fig:plots}.
A generic feature emerging from the plots is that there is always a critical number of flavors $\overline{N}_f$ for which  the normalized free energy vanishes. We interpret the change in the sign of the free energy as an indication of an instability of the system which identify with the point where large distance conformality is lost. In table \ref{fig:discrepancy} we compare  this value with the expected critical number of flavors obtained using the Ryttov-Sannino $\beta$ function
$N_f^{RS}$ (obtained for the parameter $\gamma =1$) as well as the one obtained via the ladder approximation and indicated with ${N_f^{Ladder}}$.

\begin{table}[!hptb]
\caption {Comparison of the different  critical number of flavors obtained via the $g^2$ free energy ($\overline{N}_f$), the Ryttov-Sannino $\beta$ function $N_f^{RS}$ (obtained for the parameter $\gamma =1$), and the ladder approximation ${N_f^{Ladder}}$.}
\label{fig:discrepancy}
\begin{tabular}{c | c | c | c | c || c}
$r$ 		& $N$	& $\overline{N}_f$ & $N_f^{RS}(\gamma =1)$ & discrepancy  & ${N_f^{Ladder}}$\\[1mm]
\hline\\[-3mm]
$\tiny\yng(1)$ 	& $3$ & $11.00155$ & $11$ & $0.00014$ 		& $11.914$\\[1mm]
$\tiny\yng(1)$ 	& $2$ & $7.31100$ & $7.33\ldots$ & $0.00304$ 	&$7.859$\\[1mm]
$\tiny\yng(2)$ & $3$ & $2.18463$ & $2.2$ & $0.00699$ 			&$2.502$\\[1mm]
$\tiny\yng(1,1)$ & $4$ & $7.35712$ & $7.33\ldots$ & $0.00324$	&$8.104$\\[1mm]
$G$  	& $2$	& $1.82739$ 	& $1.833\ldots$ & $0.00324$	&$2.075$
\end{tabular}
\end{table}
The agreement among these numbers is surprisingly good as it can been from the column indicated by {\it discrepancy} defined as $|\overline{N}_f - N_f^{RS}|/N_f^{RS} $. 

We still do not have a deep understanding of why these different methods agree so well among each other however we speculate that this agreement might be due to the fact that all these approaches make use of the two-loop universal coefficients of the $\beta$ function. 

Once we noticed such an agreement we asked ourselves: {\it How about supersymmetry?} We find, also in the supersymmetric case, the existence of a critical number of flavors below which the free energy changes sign. 
%
%
In table \ref{SUSYdiscrepancy} we  compare $\overline{N}_f$ with the critical number of flavors obtained using the supersymmetric all-orders $\beta$ function when setting to zero its numerator 
for both $\gamma = -1$ and $\gamma = -\frac{1}{5}$.
\begin{table}[bp]
\caption{Comparison of $\overline{N}_f$ for supersymmetric gauge theories with the critical number of flavors obtained using the supersymmetric all-orders $\beta$ function when setting to zero its numerator 
for both $\gamma = -1$ and $\gamma = -\frac{1}{5}$}
\label{SUSYdiscrepancy}
\begin{tabular}{c | c | c| c| c}
$r$ 		& $N$	& $\overline{N}_f $ & $N_f\left (\gamma =-1 (-\frac{1}{5})\right)$ & discrepancy \\[1mm]
\hline\\[-3mm]
$\tiny\yng(1)$ 	& $3$ & $7.6062$ & $4.5$ $(7.5)$ & $0.69$ $(0.014)$ \\[1mm]
$\tiny\yng(1)$ 	& $2$ & $5.1137$ & $3.0$ $(5)$ & $0.70$ $(0.023)$ \\[1mm]
$\tiny\yng(2)$ & $3$ & $1.4365$ & $0.9$ $(1.5)$ & $0.59$ $(0.042)$ \\[1mm]
$\tiny\yng(1,1)$ & $4$ & $4.9794$ & $3.0$ $(5)$ & $0.77$ $(0.004)$\\[1mm]
$G$  	& $2$	& $1.2071$ 	& $0.75$ $(1.25)$ & $0.61$ $(0.034)$
\end{tabular}
\end{table}
In this case we find a reasonable agreement when taking as critical number of flavors the one for which the anomalous dimension of the chiral superfield $\gamma$ is around $-1/5$. This is not the preferred value obtained from Seiberg's results\cite{Seiberg:1994pq} which, however, were tested via dualities only for the case of the fundamental representation.

\section{Conformal Free Energy to the Last Perturbative order}
Having at hand a perturbative expansion it is natural to go beyond the $g^2$ order. To determine the free energy at any given order, in perturbation theory, we have consistently solved for the value of the coupling evaluated  at the infrared fixed value and in the same renormalization scheme. The expressions of the fixed point value of the coupling to the highest order computed here are given in the Appendix \ref{IRFP}. 
 
 The four-loop $\beta$ function was computed in \cite{vanRitbergen:1997va} up to a normalization constant for the fourth order Casimir. In \cite{vanRitbergen:1997va} the explicit expressions for all the coefficients were given for  the fundamental and adjoint representations while we derive in Appendix \ref{sec:I4} the expressions for any totally (anti)symmetric representation for  $SU(N)$, $SO(N)$ and $Sp(N)$  gauge groups. 

\begin{figure*}[h!]
	\begin{center}
	\subfloat[$\mu = 2 \pi T$. ]{\label{subfig:g6mu2PiT}
	\includegraphics[width=0.35\textwidth]{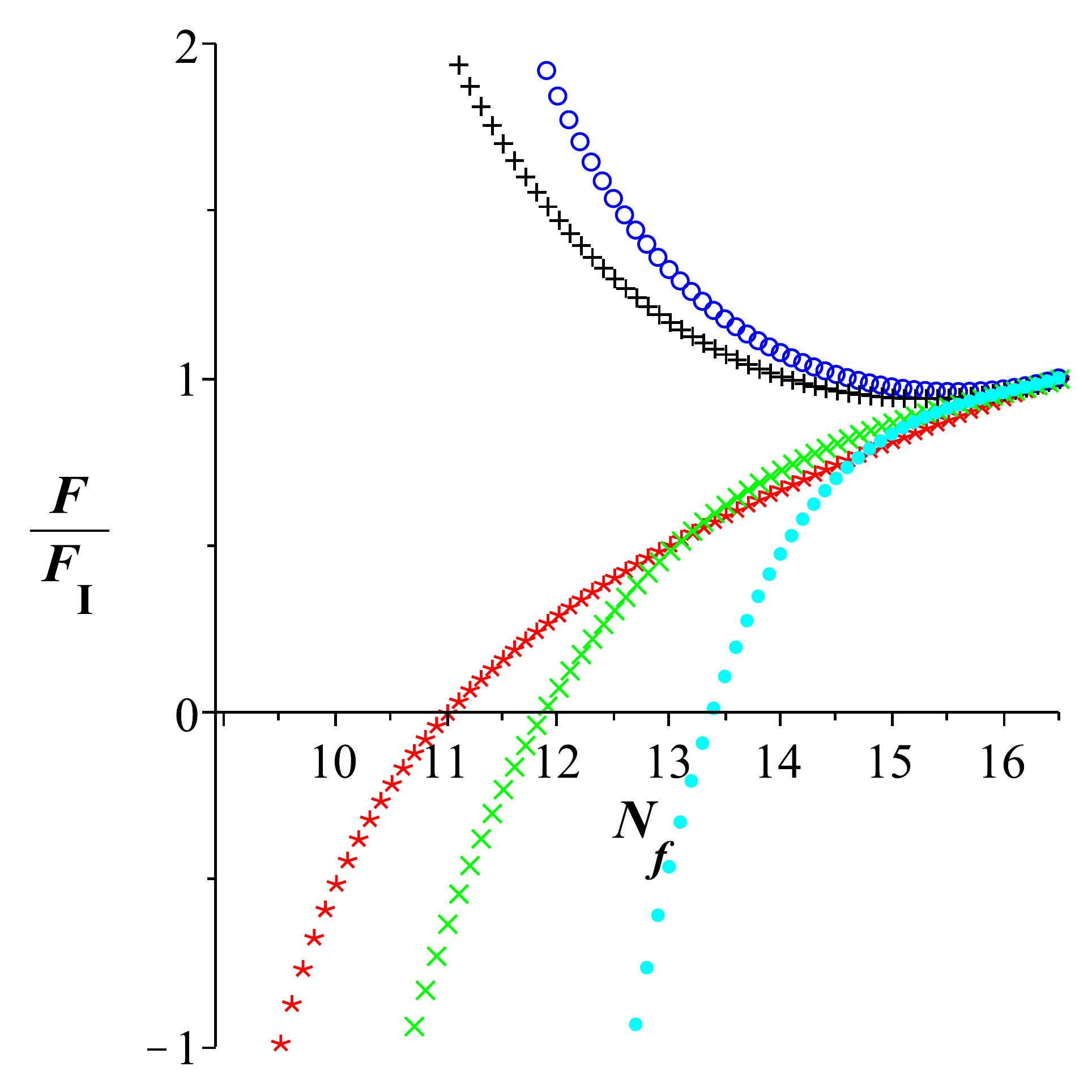}}
	\subfloat[ $\mu = g^2T$.]{\label{subfig:g6mug2T} 
	\includegraphics[width=0.15\textwidth]{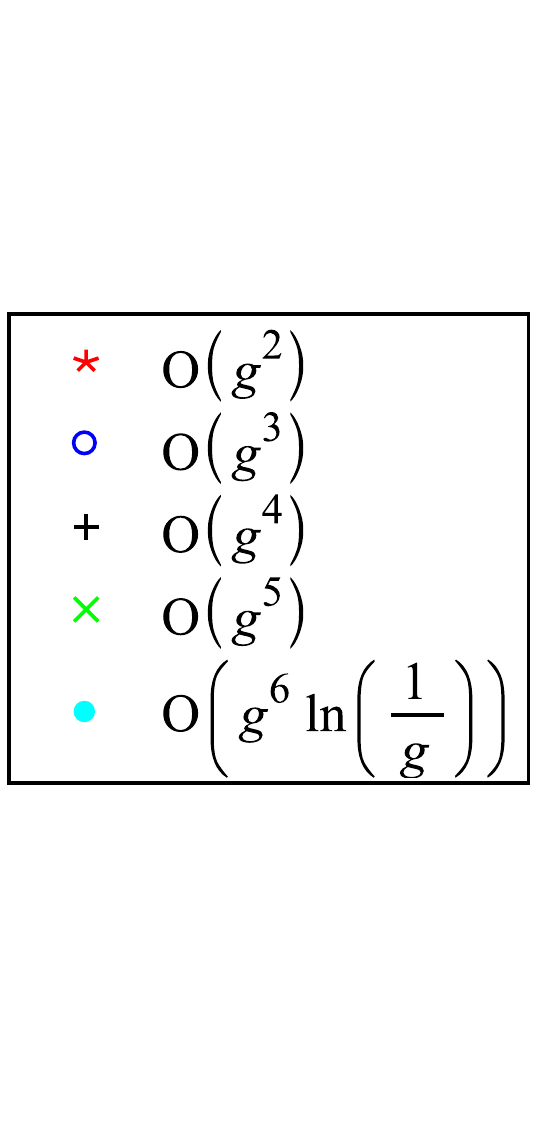}
	\includegraphics[width=0.35\textwidth]{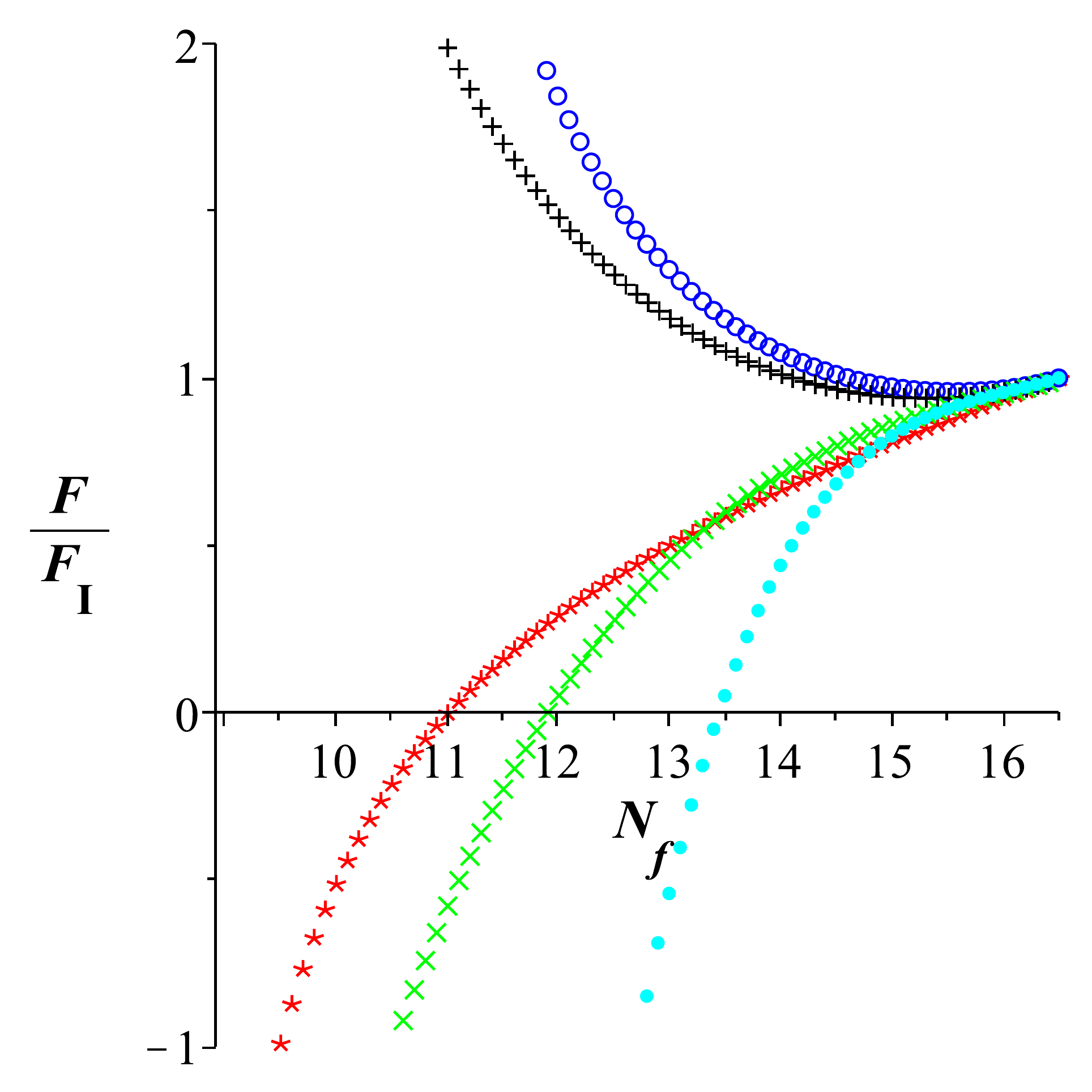}}
	\caption{Normalized free energy for different orders in $g$ computed at the
	renormalizations scale $\mu$ with
	fermions in the fundamental representation of $SU(3)$. }
	\label{fig:g6plots}
	\end{center}
\end{figure*}

Beyond the $g^3$ order one notices the emergence of logarithms of the ratio of the renormalization to the temperature scale.  Since we assumed, in our computations, the temperature scale to be such that the gauge theory coupling constant, at zero temperature, has  (quasi) reached the fixed point value it is therefore natural  to evaluate the coupling at the renormalization scale point $2\pi T$. We have, however, checked by direct evaluation of the free energy at the renormalization scale point of $g^2 T$ that, due to the logarithmic dependence, the results are rather insensitive to the choice of the reference scale as it is clear from Fig.~\ref{fig:g6plots} where we show the results for fermions in the fundamental representation for the two choices of the renormalization scale point, $2 \pi T$ (left-panel) and $g^2 T$ (right-panel).

\begin{figure*}[!hptb]
	\begin{center}
	\subfloat[$SU(2)$ with adjoint fermions. ]{\label{subfig:g6adj} 
	\includegraphics[width=0.3\textwidth]{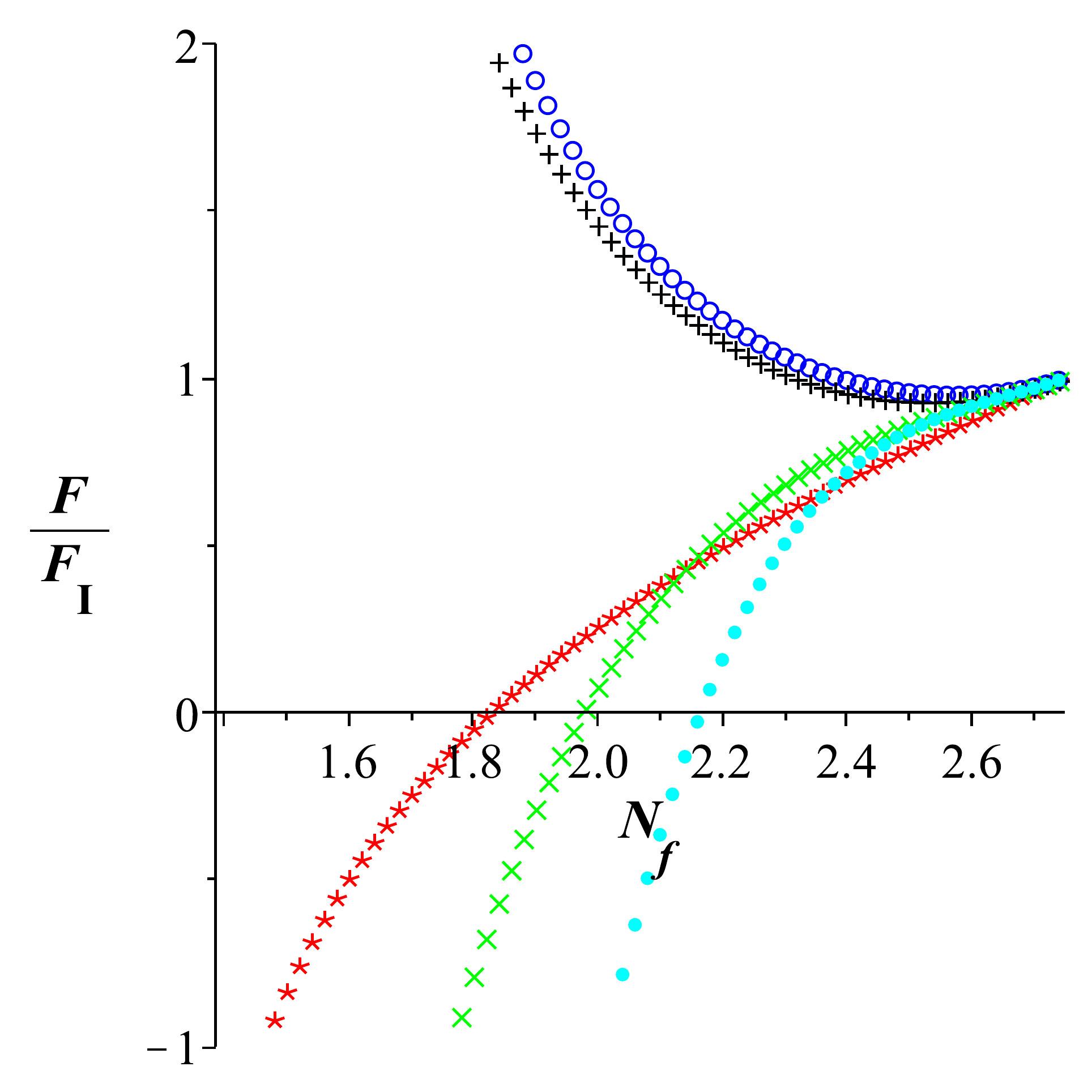}}
	\subfloat[ $SU(3)$ with two-index symmetric fermions.]{\label{subfig:g6sym} 
	\includegraphics[width=0.3\textwidth]{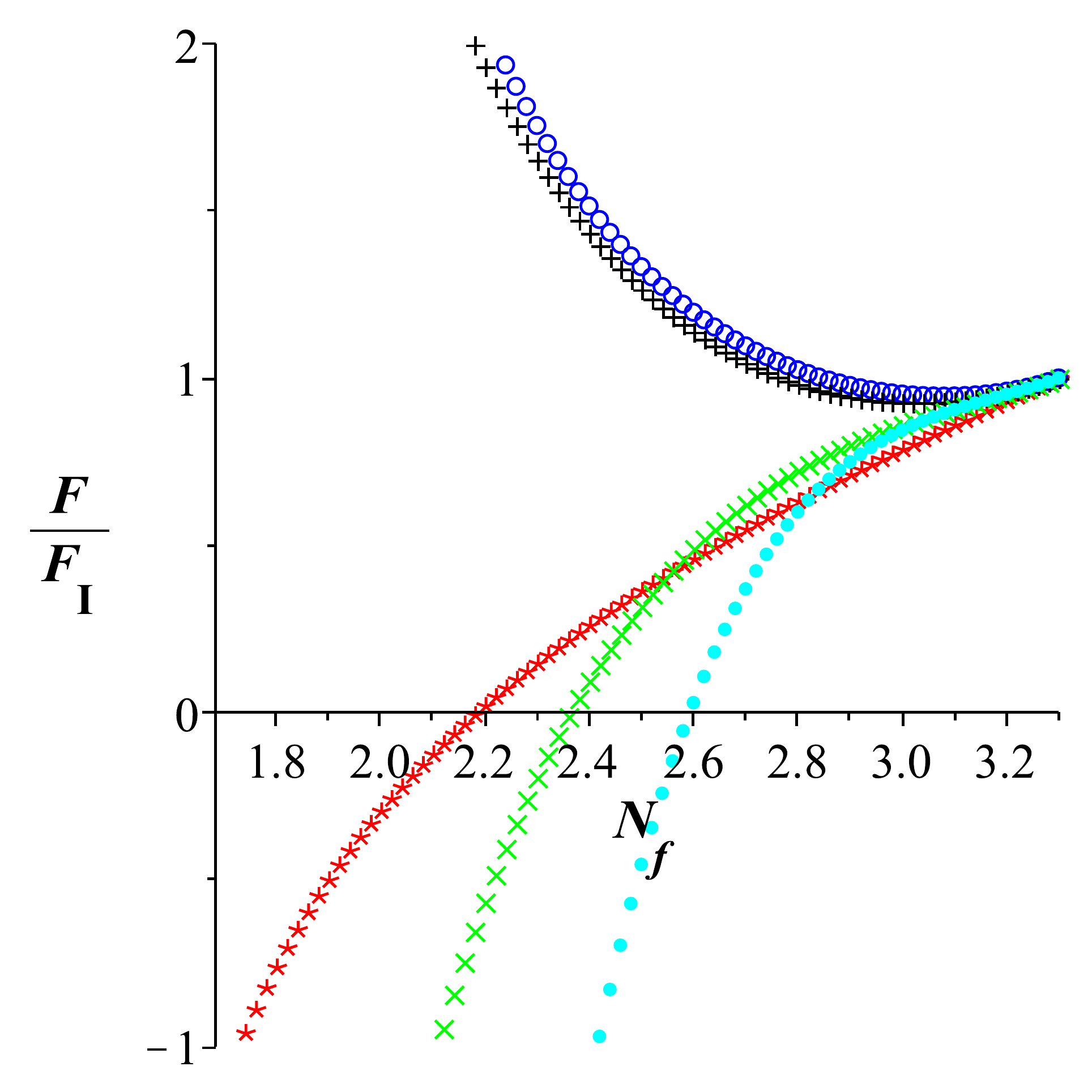}}
	\subfloat[$SU(4)$ with two-index antisymmetric fermions.]{\label{subfig:g6sntisym}
	\includegraphics[width=0.3\textwidth]{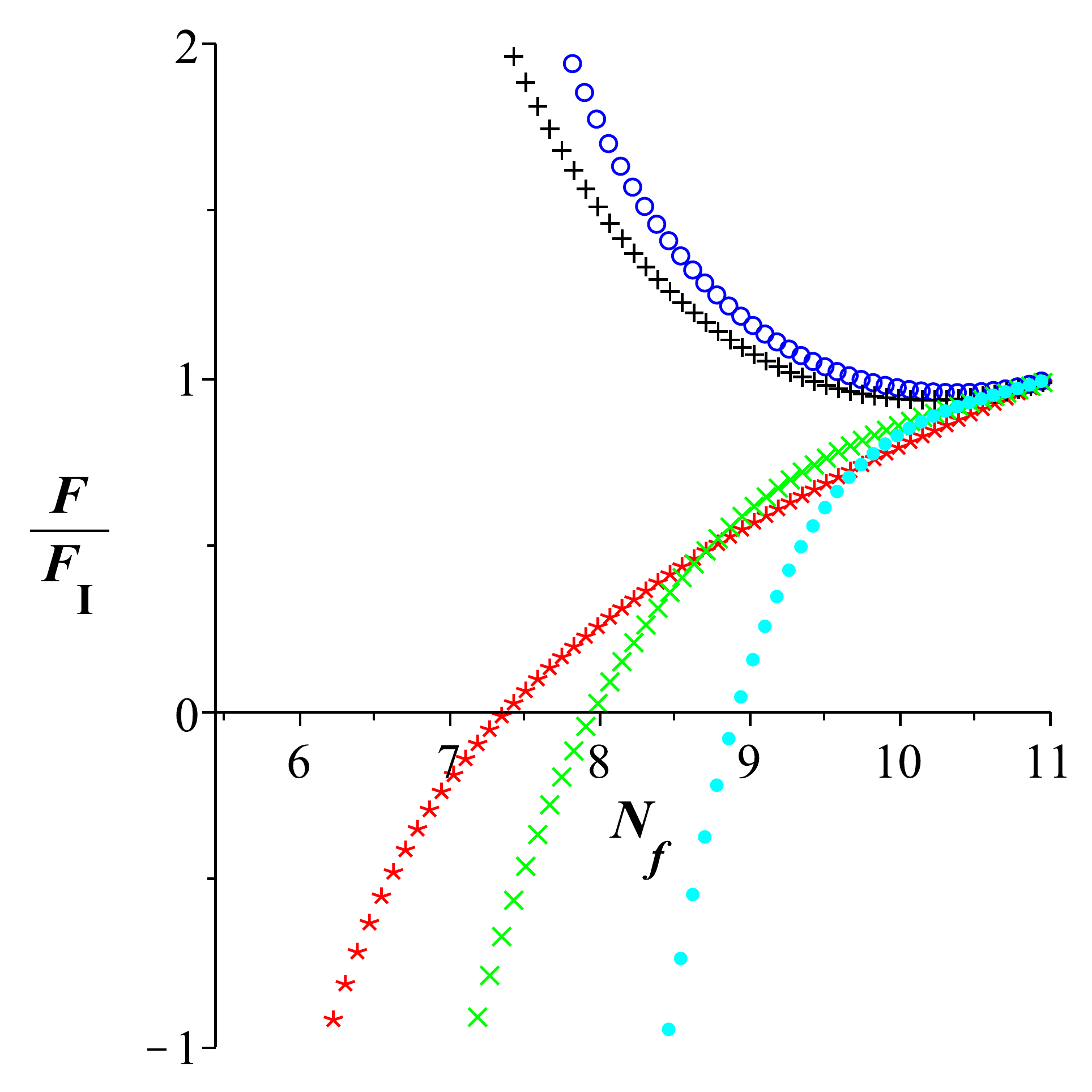}}
	\caption{Normalized free energy for different orders in $g$ computed at the
	renormalizations scale $\mu = 2 \pi T$ with
	fermions in different representation. The color-code is the same
	as in figure \ref{fig:g6plots}. }
	\label{fig:g6plots2}
	\end{center}
\end{figure*}

It is for this reason that we show in Fig.~\ref{fig:g6plots2} the results for the remaining theories evaluated at the scale $2\pi T$.  We observe the following universal behaviors:
\begin{itemize} 
\item The fee energy to the lowest interesting scheme-independent order in perturbation theory (i.e. $g^2$) changes sign at a critical number of flavors ($\overline{N}_f$),  
\item This critical value increases at the order $g^5$ and increases further at the order $g^6$,   
\item The free energy does not change sign if truncated at the order $g^3$ or $g^4$.  
\end{itemize}

\section{Conclusions}
We unveiled the finite temperature structure of gauge theories of fundamental interactions featuring a perturbative infrared  stable fixed point to the last computable order in perturbation theory. Differently from gauge theories assumed to generate a nonperturbative renormalization-invariant scale at zero temperature, like QCD, our results are perturbative in the entire energy range (i.e. for any choice of the temperature) since we can use as control parameter the number of flavors to tune the theory near the perturbative stable infrared fixed point.   

We discovered a number of universal properties, i.e. independent on the matter representation and the supersymmetric structure of the underlying gauge theory, suggesting that asymptotically free gauge theories featuring large distance conformality share very similar dynamics. 

If we were to take the point of view \cite{Kajantie:2002wa} that having exhausted the perturbative results we determined the full result for the free energy at nonzero temperature,  we would then have {\it discovered} that there is a critical number of flavors below which the free energy changes sign signaling the onset of an instability which we interpret as the end of the conformal window.

\clearpage
\clearpage
 \appendix
 \section{Banks-Zaks fixed points up to four-loops}\label{IRFP}
Here we give the exact expression for the Banks-Zaks infrared fixed point \cite{Banks:1981nn} to
different orders in $g$.

The two-loop expression is:
 \begin{align}
\tiny \frac{\alpha^*}{4\pi} =
-\frac{\beta_0}{\beta_1}
\end{align}

The three-loop expression is:
 \begin{align}
 \frac{\alpha^*}{4\pi} =
-\frac{\beta_1 + \sqrt{ \beta_1^2 - 4 \beta_2 \beta_3}}{\beta_2}.
\end{align}

The four-loop expression is:
 \begin{align}
 \frac{\alpha^*}{4\pi} =
- \frac{b_2^2 (1 + i \sqrt{3}) - b_1 (1 + i \sqrt{3}) + 2 \beta_2 b_2}{12 \beta_3 b_2},
\end{align}
where
\begin{align*}
b_1 &= 12 \beta_1 \beta_3 - 4 \beta_2^2,\\[4mm]
b_2 &= \left( 36\beta_1\beta_2\beta_3 -108\beta_0\beta_3^{2} -8\beta_2^3\frac{}{}\right.\\[2mm]
&\left. +12\sqrt{3}\sqrt{
4\beta_1^3\beta_3-\beta_1^2\beta_2^2-18
\beta_1\beta_2\beta_3\beta_0+27\beta_0^2
\beta_3^2+4\beta_0\beta_2^3}\beta_3 \right)^{\frac{1}{3}},
\end{align*}
with the $\beta_i$'s given in \cite{vanRitbergen:1997va}.

\section{Generalization for the 4th order Casimir}\label{sec:I4}
The general fourth-order Casimir invariants for
simple Lie groups were derived in \cite{Okubo:1981td,Okubo:1982dt}.
We will use the results therein to
generalize the expression for the
four-loop $\beta$ function in \cite{vanRitbergen:1997va}
to any representation of the groups
$SU(N), SO(N), Sp(N)$.
We must however keep track of the different
normalization of the Killing form in the literature,
 thus we define an overall normalization constant, $b$ as
\begin{align}
\text{Tr}f^a_{cd}f^{bcd} = b h \delta^{ab} =  \eta I_2[G] \delta^{ab},
\end{align}
where $f^{abc}$ are the structure constants and $h$ is the dual Coxeter number.
The equation defines the second-order Casimir invariant, $I_2$ with eigenvalue
$I_2[r]$ as given in \cite{Okubo:1981td} and $\eta = b/b'$ relates the
results therein to the arbitrary normalization $b$, with $b'$ chosen in \cite{Okubo:1981td, Okubo:1982dt}
to be $b' = \{2,1,2\}$ for the groups $\{SU(N),SO(N),Sp(N)\} $.
In this paper, $b$ is set to $1$. In this appendix we are following the notation introduced by Okubo \cite{Okubo:1981td} when naming the second-order Casimir invariants, i.e. $I_2[r]$, corresponding to $C_2[r]$ in the more recent literature and the one used in the main text. 

The fourth-order Casimir invariant is
related to the symmetrized fourth-order trace:
\begin{align}
 d_r^{abcd} = \frac{1}{4!} \sum_P \text{Tr} \left[  T^{a}T^{b}T^{c}T^{d} \right],
 \end{align}
 where the sum is over all permutations of the generator indices.
 We let $T^a$ be any representation of the generators for
 a simple Lie group. It is well-known that the fourth-order
 Casimir invariant is not unique, in fact the
 square of $I_2$ is also a fourth-order Casimir invariant, 
 leading to the amibiguity:
 \begin{align}
I_4' = I_4 + C (I_2)^2,
\end{align}
where $I_4'$ defines a new fourth-order Casimir invariant, with $C$ being an
arbitrary constant.
To get rid of the ambiguity one defines a \emph{modified} Casimir invariant $J_4$
with a specific metric found by Okubo and for short indicated with $\delta^{abcd}$:
\begin{align}
J_4 = \eta^2 \, \delta_{abcd} T^{a}T^{b}T^{c}T^{d}.
\end{align}
Then requiring the identity for irreducible representations:
\begin{align}
\delta_{abcd}\text{Tr}\left[ T^{a}T^{b}T^{c}T^{d}\right] = \eta^2 d[r]J_4[r],
\end{align}
where $J_4[r]$ is the eigenvalue of $J_4$ in the representation $r$,
it follows that $J_4$ satisfies similar sum-rules as $I_2$ and $I_3$,
hence it is the appropriate fourth-order Casimir invariant to work with \cite{Okubo:1981td}.
 
 It now follows that for a general representation, 
 one can write:
  \begin{align}\label{general}
 d_r^{abcd} = c_1 \delta^{abcd} + \frac{1}{3}c_2
  (\delta^{ab}\delta^{cd} + \delta^{ac}\delta^{bd} + \delta^{ad}\delta^{bc} ),
 \end{align}
 where $c_i$ are some constants dependent on the representation, $r$.
 The two terms are orthogonal.  We will only be concerned with
 the first term, as the second term was given for any representation in \cite{vanRitbergen:1997va}.
 From \cite{Okubo:1981td} one finds that
 \begin{align}
c_1 = \eta^2 \frac{d[r] J_4[r]}{d[\lambda] J_4[\lambda] (2+d[G])},
\end{align}
where $\lambda$ is 
 the defining representation, which we will take to be the fundamental one.
Then, contracting eq. \eqref{general} with $\delta^{abcd}$ one finds:
\begin{align}
\delta^{abcd}\delta_{abcd} = \frac{\eta^2 d[r] J_4[r]}{c_1} =d[\lambda] J_4[\lambda] (2+d[G]) .
\end{align}
Hence, we derive that 
\begin{align}
c_1^2 \delta^{abcd}\delta_{abcd} &= \left [ \frac{d[r] J_4[r]}{d[\lambda] J_4[\lambda]} \eta^2 \right]^2 \frac{d[\lambda] J_4[\lambda] }{(2+d[G])}\nonumber\\
			&= \left [ \frac{d[r] J_4[r]}{d[\lambda] J_4[\lambda]} b^2\right]^2 \frac{d[\lambda] J_4[\lambda] }{b'^4(2+d[G])}.
\end{align}
Writing the expression in this form, we exactly get 
the definitions of the normalization constant $\tilde{I}_4[r]$ and the
traceless tensor $d^{abcd}$ used in the four-loop $\beta$ function paper \cite{vanRitbergen:1997va}
(note that the normalization constant was defined without the tilde, but
is used here in order not to confuse it with $I_4[r]$ defined in \cite{Okubo:1981td}), i.e.
\begin{align}
\tilde{I}_4[r] &= \frac{d[r] J_4[r]}{d[\lambda] J_4[\lambda]} b^2, \\[2mm]
d^{abcd}d_{abcd} &= \frac{d[\lambda] J_4[\lambda] }{b'^4(2+d[G])}.
\end{align}
As noted in \cite{vanRitbergen:1997va}, $d^{abcd}$ is representation-independent,
and the contracted product can be written as
\begin{align}
d^{abcd}d_{abcd} [SU(N)] &= \frac{d[G] (d[G]-3)(d[G]-8)}{16\cdot 6(2+d[G])},\\[2mm]
d^{abcd}d_{abcd} [SO(N)] &= \frac{d[G] (d[G]-1)(d[G]-3)}{12(2+d[G])},\\[2mm]
d^{abcd}d_{abcd} [Sp(N)] &= \frac{d[G] (d[G]-1)(d[G]-3)}{16 \cdot 12(2+d[G])},
\end{align}
where the fundamental representation is taken as the defining representation $\lambda$,
and with
\begin{center}
\begin{tabular}{|c|c|c|c|}
\hline & & &\\[-4mm]
&  $SU(N)$ & $SO(N)$ & $Sp(N)$ \\[1mm]
\hline & & & \\[-4mm]
$d[G]$& $N^2-1$ &  $N(N-1)/2$ & $N(N+1)/2$\\[1mm]
\hline
\end{tabular}
\end{center}

Correspondingly $\tilde{I}_4[r]$ can
be derived from \cite{Okubo:1981td}, where all invariants are given.
For completion, we give the expressions here.

Denote the fundamental representations with $\{\Lambda_j\}$
corresponding to completely antisymmetric tensor representations,
while $\{k\Lambda_1 \}$ are the completely symmetric 
representations in the sense of Young's tableaux, i.e:
\begin{center}
\begin{tabular}{ccccc}
& $\tiny \young(1,2)$ & & & \\[-3mm]
$\Lambda_j \sim$ & $\tiny \vdots$& ,$\qquad$&$k\Lambda_1 \sim$ & $\tiny \young(12)\cdots\young(k)$ \\[-2mm]
& $\tiny \young(j)$ & & &
\end{tabular}
\end{center}

To simplify the expressions, we define:
\begin{align*}
\zeta_j &= N(N+1)-6j(N-j),\\
\kappa_k &= N(N-1)+6k(N+k)
\end{align*}
Then, we find for\\
\noindent \underline{$SU(N)$}:
\begin{align*}
d[\Lambda_j] &= \frac{N(N-1) \cdots (N-j+1)}{j!} \qquad 1\leq j \leq N-1\\
d[k\Lambda_1] &= \frac{N(N+1) \cdots (N+k-1)}{k!} \qquad k\geq 1\\
\tilde{I}_4[\Lambda_j] &= \frac{(N-4)!}{N!}\frac{N-j}{(j-1)!} \zeta_j \prod_{r=1}^j (N-r+1) b^2\\
\tilde{I}_4[k\Lambda_1] &= \frac{(N-1)!}{(N+3)!}\frac{N+k}{(k-1)!} \kappa_k \prod_{r=1}^k (N+r-1) b^2
\end{align*}
 
\noindent \underline{$SO(N)$}:
\begin{align*}
d[\Lambda_j] &= \frac{N!}{j!(N-j)!} \qquad 1 \leq j \leq  \frac{N-3}{2}\\[2mm]
d[k\Lambda_1] &= \frac{N+2k -2 }{k!} \frac{(N+2k-3)!}{(N-2)!} \qquad k\geq 1\\[2mm]
\tilde{I}_4[\Lambda_j] &= \frac{(N-4)!}{(j-1)! (N-j-1)!} \zeta_j b^2\\[2mm]
\tilde{I}_4[k\Lambda_1] &=\frac{(N-2+2k)(N-2+k)! }{(k-1)!(N+2)!} \times \nonumber\\
&\times \left[ N^2 - 3N + 8+6k(N-2+k)\right] b^2
\end{align*}
for $N$ odd:
\begin{align*}
d[\Lambda_\frac{N-1}{2}] = 2^\frac{N-1}{2},\qquad 
\tilde{I}_4[\Lambda_\frac{N-1}{2}] = - 2^\frac{N-9}{2}
\end{align*}
for $N$ even:
\begin{align*}
\qquad d[\Lambda_\frac{N}{2}] = 2^\frac{N-2}{2},\qquad 
\tilde{I}_4[\Lambda_\frac{N}{2}] = - 2^\frac{N-10}{2}
\end{align*}

\noindent \underline{$Sp(N)$}:
\begin{align*}
d[\Lambda_j] = &\frac{N+2-2j}{j!}\frac{(N+1)!}{(N+2-j)!} \qquad 1 \leq j \leq N/2\\[2mm]
d[k\Lambda_1] = &\frac{(N+k -1)! }{k!(N-1)!} \qquad k\geq 1\\[2mm]
\tilde{I}_4[\Lambda_j] = &\frac{(N+2-2j)(N-3)!}{(j-1)!(N-j+1)!}\times\nonumber\\
&\times\left[ N^2 + 3N + 8-6j(N+2-j)\right] b^2\\[2mm]
\tilde{I}_4[k\Lambda_1] = &\frac{(N+k)!}{(k-1)! (N+3)!} \kappa_k b^2
\end{align*}

In particular, we list the coefficients for the relevant groups in this paper, where $b$ was taken as $1$:
\begin{center}
\begin{tabular}{c | ccc }
$r\backslash \tilde{I}_4[r]$ 		& $SU(N)$	& $SO(N)$ & $Sp(N)$ \\[1mm]
\hline\\[-3mm]
$G$  	& $2Nb^2$ 	& $(N-8)b^2$ 	& $(N+8)b^2$ \\[1mm]
$\tiny\yng(1)$ 	& $b^2$ & $b^2$ & $b^2$ \\[1mm]
$\tiny\yng(2)$ & $(N+8)b^2$ & $(N+8)b^2$ & $(N+8)b^2$ \\[1mm]
$\tiny\yng(1,1)$ & $(N-8)b^2$ & $(N-8)b^2$ & $(N-8)b^2$
\end{tabular}
\end{center}
\end{document}